\documentclass[oneside,english,british]{elsart}
\usepackage[T1]{fontenc}
\usepackage[latin9]{inputenc}
\usepackage{calc}
\usepackage{units}
\usepackage{amsmath}
\usepackage{graphicx}
\usepackage{amssymb}

\makeatletter

\providecommand{\tabularnewline}{\\}

\usepackage{chemarr}
\usepackage[section,above]{placeins}
\usepackage{appendix}
\usepackage[numbers,sort&compress]{natbib}
\journal{JTB}

\@ifundefined{showcaptionsetup}{}{%
 \PassOptionsToPackage{caption=false}{subfig}}
\usepackage{subfig}
\makeatother

\usepackage{babel}

\begin{document}
\selectlanguage{english}%
\noindent \begin{frontmatter}

\begin{frontmatter}
\selectlanguage{british}%

\title{Investigating the two-moment characterisation of subcellular biochemical
networks}

\end{frontmatter}
\selectlanguage{english}%
\author[rostock]{Mukhtar Ullah},
\author[rostock,stellenbosch]{Olaf Wolkenhauer\corauthref{cor}}
\corauth[cor]{Corresponding~author.~Tel./Fax:~+49~(0)381~4987570/72\\
Emails:~\texttt{mukhtar.ullah@uni-rostock.de}~(M.U),\\
~~~\texttt{olaf.wolkenhauer@uni-rostock.de}~(O.W)\\
URL:~\texttt{www.sbi.uni-rostock.de}}
\address[rostock]{Systems Biology and Bioinformatics Group, Dept.\ of Computer Science, University of Rostock, Albert Einstein Str.\ 21, 18051 Rostock, Germany}
\address[stellenbosch]{Stellenbosch Institute for Advanced Study (STIAS), 10 Marais Street, Stellenbosch 7600, South Africa}
\selectlanguage{british}%
\begin{abstract}
While ordinary differential equations (ODEs) form the conceptual framework
for modelling many cellular processes, specific situations demand
stochastic models to capture the influence of noise. The most common
formulation of stochastic models for biochemical networks is the chemical
master equation (CME). While stochastic simulations are a practical
way to realise the CME, analytical approximations offer more insight
into the influence of noise. Towards that end, the two-moment approximation
(2MA) is a promising addition to the established analytical approaches
including the chemical Langevin equation (CLE) and the related linear
noise approximation (LNA). The 2MA approach directly tracks the mean
and (co)variance which are coupled in general. This coupling is not
obvious in CME and CLE and ignored by LNA and conventional ODE models.
We extend previous derivations of 2MA by allowing a) non-elementary
reactions and b) relative concentrations. Often, several elementary
reactions are approximated by a single step. Furthermore, practical
situations often require the use relative concentrations. We investigate
the applicability of the 2MA approach to the well established fission
yeast cell cycle model. Our analytical model reproduces the clustering
of cycle times observed in experiments. This is explained through
multiple resettings of MPF, caused by the coupling between mean and
(co)variance, near the G2/M transition.
\end{abstract}
\selectlanguage{english}%
\end{frontmatter}

\selectlanguage{british}%

\section{Introduction}

At a coarse level, cellular functions are largely determined by spatio-temporal
changes in the abundance of molecular components. At a finer level,
cellular events are triggered by discrete and random encounters of
molecules \cite{paulsson2006}. This suggests a deterministic modelling
approach at the coarse level (cell function) and a stochastic one
at the finer level (gene regulation) \cite{rao2002,paulsson2004,kaern2005,raser2005,pedraza2005,mantzaris2007,ansel2008,lipniacki2006,paszek2007,becskei2005}.
However, stochastic modelling is necessary when noise propagation
from processes at the fine level changes cellular behaviour at the
coarse level.

Stochasticity is not limited to low copy numbers. The binding and
dissociation events during transcription initiation are the result
of random encounters between molecules \cite{kaern2005}. If molecules
are present in large numbers and the molecular events occur frequently,
the randomness would cancel out (both within a single cell and from
cell to cell) and the average cellular behaviour could be described
by a deterministic model. However, many subcellular processes, including
gene expression, are characterised by infrequent (rare) molecular
events involving small copy numbers of molecules \cite{kaern2005,paulsson2006}.
Most proteins in metabolic pathways and signalling networks, realising
cell functions, are present in the range 10-1000 copies per cell \cite{berg2000,levine2007a,paulsson2005}.
For such moderate/large copy numbers, noise can be significant when
the system dynamics are driven towards critical points in cellular
systems which operate far from equilibrium \cite{elf2003a,tao2005,zhang2006}.
The significance of noise in such systems has been demonstrated for
microtubule formation \cite{dogterom1993}, ultrasensitive modification
and demodification reactions \cite{berg2000}, plasmid copy number
control \cite{paulsson2001}, limit cycle attractor \cite{qian2002},
noise-induced oscillations near a macroscopic Hopf bifurcation \cite{vilar2002},
and intracellular metabolite concentrations \cite{elf2003c}. 

Noise has a role at all levels of cell function. Noise, when undesired,
may be suppressed by the network (e.g. through negative feedback)
for robust behaviour \cite{samad2004a,thattai2002,fraser2004a,morishita2004,rao2002,paulsson2000a}.
However, all noise may not be rejected and some noise may even be
amplified from process to process, and ultimately influencing the
phenotypic behaviour of the cell \cite{hornung2008,lan2006,pedraza2005,becskei2005,shibata2008}.
Noise may even be exploited by the network to generate desired variability
(phenotypic and cell-type diversification) \cite{blomberg2006,chen2006,hasty2000,rao2002,yoda2007}.
Noise from gene expression can induce new dynamics including amplification
(stochastic focusing) \cite{paulsson2000,samoilov2005,pedraza2005},
bistability (switching between states) and oscillations \cite{ferrell2001,aumaitre2007,ozbudak2004,artyomov2007},
that is both quantitatively and qualitatively different from what
is predicted or possible deterministically.

The most common formulation of stochastic models for biochemical networks
is the chemical master equation (CME). While stochastic simulations
\cite{turner2004} are a practical way to realise the CME, analytical
approximations offer more insights into the influence of noise on
cell function. Formally, the CME is a continuous-time discrete-state
Markov process \cite{singer1953,gillespie1977,kampen2007}. For gaining
intuitive insight and a quick characterisation of fluctuations in
biochemical networks, the CME is usually approximated analytically
in different ways \cite{kampen2007,goutsias2006}, including the frequently
used the chemical Langevin approach \cite{gillespie2000,kampen2007h,steuer2004,zamborszky2007},
the linear noise approximation (LNA) \cite{elf2003a,hayot2004,scott2005,scott2006}
and the two-moment approximation (2MA) \cite{goutsias2007,gomez-uribe2007,ferm2007a}.

Of the analytical approaches mentioned above, we here focus on the
2MA approach because of its representation of the coupling between
the mean and (co)variance. The traditional Langevin approach is based
on the assumption that the time-rate of abundance (copy number or
concentration) or the flux of a component can be decomposed into a
deterministic flux and a Langevin noise term, which is a Gaussian
(white noise) process with zero mean and amplitude determined by the
the dynamics of the system. This separation of noise from the system
dynamics may be a reasonable assumption for \emph{external noise}
that arises from the interaction of the system with other systems
(like the environment), but cannot be assumed for internal noise that
arises from within the system \cite{kaern2005,raser2005,paulsson2005,becskei2005,dublanche2006,shahrezaei2008}.
As categorically discussed in \cite{kampen2007h}, internal noise
is not something that can be isolated from the system because it results
from the discrete nature of the underlying molecular events. Any noise
term in the model must be derived from the system dynamics and cannot
be presupposed in an \emph{ad hoc} manner. However the chemical Langevin
equation (CLE) does not suffer from the above criticism because Gillespie
\cite{gillespie2000} derived it from the CME description. The CLE
allows much faster simulations compared to the exact stochastic simulation
algorithm (SSA) \cite{gillespie1977} and its variants. The CLE is
a stochastic differential equation (dealing directly with random variables
rather than moments) and has no direct way of representing the mean
and (co)variance and the coupling between the two. That does not imply
that CLE ignores the coupling like the LNA which has the same mean
as the solution of the deterministic model.

The merits of the 2MA compared to alternative approximations have
been discussed in \cite{gomez-uribe2007,goutsias2007,tang2008}. In
\cite{ferm2007a}, the 2MA is developed as an approximation of the
master equation for a generic Markov process. In \cite{gomez-uribe2007},
the 2MA framework is developed under the name {}``mass fluctuation
kinetics'' for biochemical networks composed of elementary reactions.
The authors demonstrate that the 2MA can reveal new behaviour like
stochastic focusing and bistability. Another instance of the 2MA is
proposed in \cite{goutsias2006,goutsias2007} under the names {}``mean-field
approximation'' and {}``statistical chemical kinetics''. Again,
the authors assume elementary reactions so that the propensity function
is at most quadratic in concentrations. The authors evaluate the accuracy
of the 2MA against the alternatives (such as LNA) for a few toy models.
The derivation of the 2-MA for more general systems with non-elementary
reactions is one motivation for the present paper.

The 2MA approaches referred to above assume absolute concentrations
(copy number divided by some fixed system size parameter). In systems
biology, however, models often use relative concentrations that have
arbitrary units \cite{novak2001,novak2005,tyson2002,csikasz-nagy2006}.
In general, the concentration of each component in the system may
have been obtained by a different scaling parameter, rather than using
a global system size. For such models, the above mentioned approaches
need modification. This was another motivation for our derivation
in this paper.

In the present paper we develop a compact derivation of the first
two-moments, the mean and (co)variance of the continuous-time discrete-state
Markov process that models a biochemical reaction system by the CME.
This derivation is an extension of previous derivations, taking into
account arbitrary concentrations and non-elementary reactions. The
matrix form of our derivation allows for an easy interpretation. Using
these analytical results, we develop our \foreignlanguage{english}{2MA}
model of the fission yeast cell cycle which has two sets of ODEs:
one set for the mean protein concentrations and the other set for
concentration (co)variances. Numerical simulations of our model show
a considerably different behaviour. Especially, for the \emph{wee1$^{\text{-}}$
cdc25$\Delta$} mutant (hereafter referred simply as double-mutant),
the timings of S-phase and M-phase are visibly different from those
obtained for a deterministic model because of the oscillatory behaviour
of the key regulator. Since the 2MA is only an approximation, we investigate
its validity by comparing the statistics computed from the 2MA model
with experimental data.

The rest of this paper is organised as follows. In the first section
we introduce the basic terminology and notation. Then the system of
ODEs forming the 2MA approach is presented. Next, we introduce an
application to the fission yeast cell cycle model \cite{novak2001}.
We present a 2MA model of the cell cycle, followed by a comparison
to the experimental data and conclusions. The appendices contain full
derivations of the 2MA model, further proofs and additional tables.

\section{Stochastic modelling of biochemical systems}

Imagine a well-mixed homogeneous cellular compartment of a fixed volume
$V$ at thermal equilibrium that contains molecules of $s$ different
kinds (each kind referred to as a chemical\emph{ component} or \emph{species})
interacting in $r$ distinct ways (each way referred to as a reaction\emph{
channel} or \emph{step}). Since these biochemical reactions occur
by random encounters of reactant molecules, the copy number of a particular
component present in the system at time $t$ fluctuates. The state
of the cellular system is described by the $s\times1$ random vector
$N(t)$ whose $i$th element is the copy number $N_{i}(t)$ of the
$i$th species present in the system at time $t$. Each (time-varying)
element $N_{i}(t)$ is a stochastic process, where $N_{i}(t)=n_{i}$
means that $n_{i}$ molecules of the $i$th species are present in
the system at time $t$. The $s\times1$ vector $n$, with elements
$n_{i}$, is thus a sample (or a value) of the stochastic process
$N(t)$. The stochastic process is characterised by the (time-dependent)
probability distribution $P(n,t)$, that is the probability of $N(t)=n$
given a fixed initial condition $N(0)=n^{0}$. The probability distribution
itself is characterised by its moments.

We can describe the system state at time $t$ by the $s\times1$ vector
$X(t)$ whose $i$th element is the concentration $X_{i}(t)$ of the
$i$th component. The concentration $X_{i}(t)$ is, in general, the
copy number $N_{i}(t)$ divided by some fixed scaling parameter $\mathit{\Omega}_{i}$
specific to that component. In other words\[
N_{i}(t)=\mathit{\Omega}_{i}X_{i}(t),\quad n_{i}=\mathit{\Omega}_{i}x_{i}\,.\]
Each concentration $X_{i}(t)$ is a stochastic process, where $X_{i}(t)=x_{i}$
means that the concentration of the $i$th component at time $t$
is $x_{i}$. The $s\times1$ vector $x$, with elements $x_{i}$,
is thus a sample of the stochastic process $X(t)$. The copy number
and concentration (vectors) are related by\[
N(t)=\mathit{\Omega}X(t),\quad n=\mathit{\Omega}x,\]
where $\mathit{\Omega}$ is the diagonal matrix with $\mathit{\Omega}_{i}$
being its $i$th diagonal element. 

Commonly, all components are scaled by a single parameter, in which
case $\mathit{\Omega}$ is a scalar known as the \emph{system size.}
A common choice for the system size is some multiple of the volume
$V$ of the system. For molar concentrations, the system size chosen
is $\mathit{\Omega}=N_{A}V$ where $N_{A}$ is the Avogadro's constant.
In systems biology, one often uses relative concentrations $x_{i}$
where $\mathit{\Omega}_{i}$ is some fixed copy number specific to
component $i$. The simplest case of relative concentrations uses
a single (maximum) copy number $n_{\max}$ for all components. Note
that our approach is developed for the general case which allows for
relative concentrations instead of assuming one global system-size
$\mathit{\Omega}$ as done in \cite{yi2008,goutsias2007,gomez-uribe2007,scott2005,tao2005}.

If we assume that the molecules are well mixed and are available everywhere
for a reaction (space can be ignored), then the probability of a reaction
in a short time interval depends almost entirely on the most recent
copy numbers (and not its earlier values). In other words, the stochastic
process $N(t)$ of copy numbers is \emph{Markovian} in continuous-time.
Since changes in the copy numbers require the occurrences of reactions
which are discrete event phenomena, $N(t)$ is referred as a \emph{jump
process}. The Markov property implies that each reaction channel $j$
can be characterised by a \emph{reaction propensity} $a_{j}(n)$ defined
such that, in state $n$, the probability of one occurrence of reaction
channel $j$ in a vanishingly short time interval of length $dt$
is $a_{j}(n)dt$.

The transition from state $n$ to the state determined by the $j$th
reaction will be represented by the following scheme\[
n\xrightarrow{\quad a_{j}(n)\quad}n+S_{\centerdot j}\]
where $S_{\centerdot j}$ is the $j$th column of the \emph{stoichiometry
matrix} $S$ whose element $S_{ij}$ denotes the change in copy number
of the $i$th component resulting from the occurrence of the $j$th
channel. Similarly the transitions towards state $n$ from the state
determined by the $j$th reaction can be represented by\[
n-S_{\centerdot j}\xrightarrow{\quad a_{j}(n-S_{\centerdot j})\quad}n\]
where the argument of the propensity function $a_{j}$ is $n-S_{\centerdot j}$
which is the assumed current state. Transitions away from state $n$
will decrease the probability $P(n,t)$ while those towards state
$n$ will increase it. Since this is equally true for each reaction
channel, during a short time interval of length $\Delta t$, the change
in the probability is given by\[
P(n,t+\Delta t)-P(n,t)=\sum_{j=1}^{r}P(n-S_{\centerdot j},t)a_{j}(n-S_{\centerdot j})\Delta t-\sum_{j=1}^{r}P(n,t)a_{j}(n)\Delta t+o(\Delta t)\]
where $o(\Delta t)$ represents terms that vanish faster than $\Delta t$
as the later approaches zero. As $\Delta t$ approaches zero in the
above system of equations, we are led to what is known as the \emph{chemical
master equation} (CME): \begin{equation}
\frac{d}{dt}P(n,t)=\sum_{j=1}^{r}\biggl[a_{j}(n-S_{\centerdot j})P(n-S_{\centerdot j},t)-a_{j}(n)P(n,t)\biggr]\,.\label{eq:cme}\end{equation}
We will switch between the two alternative notations $\frac{d}{dt}\phi(t)$
and $\tfrac{d\phi}{dt}$ for any scalar quantity $\phi(t)$. We will
prefer the later when dependence on time variable is implicitly clear. 

Since there is one equation for each state $n$ and there is potentially
a large number of possible states, it is impractical to solve the
CME. In most cases, we are interested in the first two-moments: component-wise
copy number means\[
\mathrm{E}\left[N_{i}(t)\right]=\sum_{n}n_{i}P(n,t),\]
and the covariances \[
\mathrm{Cov}\left(N_{i}(t),N_{k}(t)\right)=\mathrm{E}\Bigl[\bigl(N_{i}(t)-\mathrm{E}\left[N_{i}(t)\right]\bigr)\bigl(N_{k}(t)-\mathrm{E}\left[N_{k}(t)\right]\bigr)\Bigr],\]
between copy numbers of component pairs. These covariances form the
covariance matrix in which the diagonal elements are component-wise
variances. 

In the present paper, we are interested in the mean concentration
vector $\mu(t)$ with elements \[
\mu_{i}(t)=\mathrm{E}\left[X_{i}(t)\right]=\frac{\mathrm{E}\left[N_{i}(t)\right]}{\mathit{\Omega}_{i}}\]
and the concentration covariance matrix $\sigma(t)$ with elements\[
\sigma_{ik}(t)=\mathrm{Cov}\left(X_{i}(t),X_{k}(t)\right)=\frac{\mathrm{Cov}\left(N_{i}(t),N_{k}(t)\right)}{\mathit{\Omega}_{i}\mathit{\Omega}_{k}}\]
Hereafter, we leave out the dependence on time to simplify the notation,
but include it occasionally when causing confusion. 

\selectlanguage{english}%

\subsection{Continuous approximations of the jump process $N(t)$}

While the stochastic simulation algorithm and extensions provide a
way to generate sample paths of copy numbers for a biochemical system,
the need for repeating many simulation runs to get an idea of the
probability distribution in terms of its moments (mean and (co)variance)
become increasing time consuming and even impractical for larger systems.
Therefore attempts have been made towards approximations of the CME,
the most notable being the chemical Langevin equation (CLE) by Gillespie~\cite{gillespie2000}.
He obtained that continuous approximation for the incremental change
in copy number during a short interval $[t,t+dt]$ where the interval
length $dt$ \foreignlanguage{british}{satisfies two conditions: (i)
It is small enough that the propensity does not change {}``appreciably''
during the interval, and (ii) is large enough that the expected number
of occurrences $\mathrm{E}\left[Z_{j}(t+dt)-Z_{j}(t)\right]$ of each
reaction channel $j$ during the interval is much larger than unity.
That }continuous\foreignlanguage{british}{ approximation takes the
form of the CLE\begin{equation}
N_{i}^{c}(t+dt)-N_{i}^{c}(t)=\sum_{j=1}^{r}S_{ij}a_{j}\left(N^{c}(t)\right)dt+\sum_{j=1}^{r}S_{ij}\sqrt{a_{j}\left(N^{c}(t)\right)dt}\,\mathcal{N}_{j}(t)\,.\label{eq:cle}\end{equation}
 Here $N^{c}(t)$ denotes the continuous Markov process approximating
the jump process $N(t)$, and the set $\left\{ \mathcal{N}_{j}(t)\right\} $
are statistically independent Gaussian random variables each with
zero mean and unit variance. The probability density function $P^{c}(n,t)$
of the continuous Markov process $N^{c}(t)$ obeys the (forward) Fokker-Planck
equation (FPE)~\cite{gillespie1996,gillespie2000}\begin{equation}
\frac{\partial}{\partial t}P^{c}(n,t)=\sum_{j=1}^{r}\Bigl(-\sum_{i=1}^{s}S_{ij}\frac{\partial}{\partial n_{i}}+\frac{1}{2}\sum_{i,k=1}^{s}S_{ij}S_{kj}\frac{\partial^{2}}{\partial n_{i}\partial n_{k}}\Bigr)\bigl[a_{j}(n)P^{c}(n,t)\bigr]\,.\label{eq:fpe}\end{equation}
In effect, condition (i) allows a Poissonian approximation of $Z_{j}(t+dt)-Z_{j}(t)$
and condition (ii) allows a normal approximation of the Poissonian.
The two conditions seem conflicting and require the existence of a
domain of macroscopically infinitesimal time intervals. Although the
existence of a such a domain cannot be guaranteed, Gillespie argues
that this can be found for most practical cases. Admitting that, {}``it
may not be easy to continually monitor the system to ensure that conditions
(i) and (ii) {[}..{]} are satisfied.'' He justifies his argument
by saying that this {}``will not be the first time that Nature has
proved to be unaccommodating to our purposes.'' }\cite{gillespie2000}.

\selectlanguage{british}%
Generating sample paths of \eqref{eq:cle} is orders of magnitude
faster than doing the same for the CME because it essentially needs
generation of normal random numbers. See \cite{higham2001} for numerical
simulation methods of stochastic differential equations such as \eqref{eq:cle}.
However, solving the nonlinear FPE \eqref{eq:fpe} for the probability
density is as difficult as the CME. Therefore, on the analytical side,
the CLE and the associated nonlinear FPE do not provide any significant
advantage. That leads to a further simplification referred to as the
\emph{linear noise approximation} (LNA) \cite{goutsias2006,kampen2007}.
The LNA is a linear approximation of the nonlinear FPE \eqref{eq:fpe}
obtained by linearising the propensity function around the mean. The
solution of the LNA is a Gaussian distribution with a mean that is
equal to the solution of the deterministic ODE model and a covariance
matrix that obeys a linear ODE. This is the main drawback of LNA because,
for system containing at least one biomolecular reactions, the mean
of a stochastic model is not equal to the solution of deterministic
ODEs, as shown next.

\subsection{Mean of the stochastic model}

The mean copy number for the $i$th component obeys the ODE \begin{equation}
\frac{d}{dt}\mathrm{E}\left[N_{i}(t)\right]=\sum_{j=1}^{r}S_{ij}\mathrm{E}\left[a_{j}\bigl(N(t)\bigr)\right]\label{eq:mean-copynumber}\end{equation}
which is derived in Appendix A1. In general, the expectation on the
right of \eqref{eq:mean-copynumber} involves involves the unknown
probability distribution $P(n,t)$. In other words, the mean copy
number depends not just on the mean itself, but also involves higher-order
moments, and therefore \eqref{eq:mean-copynumber} is, in general,
not closed in the mean unless the reaction propensity is a linear
function of $N$ which is the case only for zero- and first-order
reactions. Take the example of a first-order reaction $X\xrightarrow{k}Y$
with $n$ denoting the copy number of its reactant and $k$ denoting
the reaction coefficient. The reaction propensity $a(n)=kn$ (mass
action kinetics) is linear in $n$. From probability theory, the expectation
becomes $\mathrm{E}(kN)=k\mathrm{E}(N)$ and thus we do not need to
know the probability distribution for solving the ODE in the mean.
Only if \emph{all} reactions \emph{elementary} and are of zero or
first-order, we have exact equations for the evolution of mean:\[
\frac{d}{dt}\mathrm{E}\left[N_{i}(t)\right]=\sum_{j=1}^{r}S_{ij}a_{j}\bigl(\mathrm{E}\left[N(t)\right]\bigr)\]
which corresponds to the ODE system for the deterministic model which
treats the copy numbers $n(t)$ as a continuous time-varying quantity
that can be uniquely predicted for a given initial condition. For
systems containing second (and higher) order reactions, $a(n)$ is
a nonlinear function and the evolution of the mean cannot be determined
by the mean alone. Instead the mean depends on higher-order moments,
and hence the deterministic ODE model and the LNA cannot be used to
describe the mean in \eqref{eq:mean-copynumber}.

\subsection{The 2MA approach}

The present section provides only a brief outline of the 2MA approach
and we refer to the Appendix A1 for a detailed derivation.

An exact and closed representation of mean is not possible in general,
as evident from \eqref{eq:mean-copynumber}. The same is true for
(co)variance and higher-order moments. One way to solve this problem
is by repeating many stochastic simulation runs based on CME or the
CLE, and computing the desired moments from the ensemble runs. An
alternative is to find approximations to the exact ODEs such as \eqref{eq:mean-copynumber}
for the moments. The 2MA is one such attempt which assumes closure
to the first two-moments: the mean and (co)variance. A scheme of chemical
reactions or a system of deterministic ODEs is the starting point.
From this are concluded the reaction propensities $a_{j}(n)$ which
appear as coefficients in the CME describing the time derivative of
the probability distribution $P(n,t)$. By taking the first two-moments
of the CME and subsequent simplifications followed by appropriate
scaling, two sets of ODEs for the mean concentration vector $\mu(t)$
and covariance matrix $\sigma(t)$ are derived. This is followed by
Taylor expansions of any nonlinear functions involving the propensity
vector $a(n)$. Ignoring central moments of 3rd and order higher eventually
leads to the 2MA system: \begin{align}
\frac{d\mu}{dt} & =f(\mu)+\varepsilon_{f}(\mu,\sigma)\label{eq:2MA-mu}\\
\frac{d\sigma}{dt} & =A(\mu)\sigma+\sigma A(\mu)^{T}+\mathit{\Omega}^{-\nicefrac{1}{2}}\left[B(\mu)+\varepsilon_{B}(\mu,\sigma)\right]\left(\mathit{\Omega}^{-\nicefrac{1}{2}}\right)^{T}\label{eq:2MA-sigma}\end{align}
where the superscript $T$ denotes transpose of a matrix and\begin{equation}
\begin{aligned}f_{i}(x) & =\frac{1}{\mathit{\Omega}_{i}}\sum_{j=1}^{r}S_{ij}a_{j}(\mathit{\Omega}x)\\
\varepsilon_{f_{i}}(\mu,\sigma) & =\frac{1}{2}\sum_{k,l}\left[\frac{\partial^{2}f_{i}}{\partial x_{k}\partial x_{l}}\right]_{x=\mu}\sigma_{kl}\\
A_{ik}(x) & =\frac{\partial f_{i}(x)}{\partial x_{k}}\\
B_{ik}(x) & =\frac{1}{\sqrt{\mathit{\Omega}_{i}\mathit{\Omega}_{k}}}\sum_{j=1}^{r}S_{ij}S_{kj}a_{j}(\mathit{\Omega}x)\\
\varepsilon_{B_{ik}}(\mu,\sigma) & =\frac{1}{2}\sum_{i',k'}\left[\frac{\partial^{2}B_{ik}}{\partial x_{i'}\partial x_{k'}}\right]_{x=\mu}\sigma_{i'k'}\,.\end{aligned}
\label{eq:2MA-aux}\end{equation}
 The derivation of these equations is given in Appendix A1. The \emph{effective
flux} on the right in \eqref{eq:2MA-mu} is the sum of a \emph{deterministic
flux} $f(\mu)$ and a \emph{stochastic flux} $\varepsilon_{f}(\mu,\sigma)$,
the latter determined by the dynamics of both the mean and (co)variance.
This influence of the (co)variance implies that knowledge of fluctuations
is important for a correct description of the mean. This also indicates
an advantage of the stochastic framework over its deterministic counterpart:
starting from the same assumptions and approximations, the stochastic
framework allows us to describe the influence of fluctuations on the
mean. This can be posed as the central phenomenological argument for
stochastic modelling.

Note that \eqref{eq:2MA-mu} is exact for systems where no reaction
has an order higher than two because then 3rd and higher derivatives
of propensity are zero. In \eqref{eq:2MA-sigma}, the \emph{drift
matrix} $A(\mu)$ reflects the noise dynamics for relaxation to the
steady state and the (Taylor approximation to the 2nd order of) \emph{diffusion
matrix} $B(n)$ the randomness (fluctuation) of the individual events.
The scaling by $\mathit{\Omega}$ confirms the inverse relationship
between the noise, as measured by (co)variance, and the system size.
Note the influence of the mean on the (co)variance in \eqref{eq:2MA-sigma}.

A deterministic model treats concentrations $x(t)$ as continuous
variables that can be predicted entirely from the initial conditions.
Hence there is no noise term in the deterministic model and the ODEs
reduce to $\dot{x}=f(x)$.

Since the 2MA approach is based on the truncation of terms containing
3rd and higher-order moments, any conclusion from the solution of
2MA must be drawn with care. Ideally, the 2MA should be complemented
and checked with a reasonable number of SSA runs.

In \cite{gomez-uribe2007,goutsias2007}, the 2MA has been applied
biochemical systems, demonstrating quantitative and qualitative differences
between the mean of the stochastic model and the solution of the deterministic
model. The examples used in \cite{gomez-uribe2007,goutsias2007} all
assume elementary reactions (and hence propensities at most quadratic)
and the usual interpretation of concentration as the moles per unit
volume. In the next section, we investigate the 2MA for complex systems
with non-elementary and relative concentrations. The reason for our
interest in non-elementary reactions is the frequent occurrence of
rational propensities (reaction rates), e.g. Michaelis-Menten type
and Hill type, in models in the system biology literature (e.g.~\cite{tyson2003}).

\section{Fission yeast cell cycle modelling}

The growth and reproduction of organisms requires a precisely controlled
sequence of events known as the cell cycle \cite{morgan2007}. On
a coarse scale, the cell cycle is composed of four phases: the replication
of DNA (S phase), the separation of DNA (mitosis, M phase), and the
intervening phases (gapes G1 and G2) which allow for preparation,
regulation and control of cell division. The central molecular components
of cell cycle control system have been identified \cite{nurse2000,morgan2007}.

Cell cycle experiments show that cycle times (CTs) have different
patterns for the wild type and for various mutants \cite{sveiczer1996,sveiczer2002}.
For the wild type, the CTs have more or less a constant value near
150 min ensured by a size control mechanism: mitosis happens only
when the cell has reached a critical size. The value 150 min has been
considered in \cite{sveiczer2000,sveiczer2002,steuer2004,yi2008}
as the CT of an average WT cell (also referred to as the {}``mass-doubling
time''). The double-mutant of fission yeast (namely \emph{wee1$^{\text{-}}$
cdc25}$\Delta$) exhibits quantised cycle times: the CTs get clustered
into three different groups (with mean CTs of 90, 160 and 230 min).
The proposed explanation for the quantised cycle times is a weakend
positive feedback loop (due to wee1 and cdc25) which means cells reset
(more than once) back to G2 from early stages of mitosis by premature
activation of a negative feedback loop \cite{sveiczer2000,sveiczer2002}.

Many deterministic ODE models describing the cell cycle dynamics have
been constructed \cite{novak1998,novak2001,novak2003,tyson2002}.
These models can explain many aspects of the cell cycle including
the size control for both the wild type and mutants. Since deterministic
models describe the behaviour of a non-existing \textquoteleft{}average
cell\textquoteright{}, neglecting the differences among cells in culture,
they fail to explain curious behaviours such as the quantised cycle
times in the double-mutant. To account for such curiosities in experiments,
two stochastic models were constructed by Sveiczer: The first model
\cite{sveiczer2000,sveiczer2002} introduces (external) noise into
the rate parameter of the protein Pyp3. The second model \cite{sveiczer2001}
introduces noise into two cell and nuclear sizes after division asymmetry.
Full stochastic models that treat all the time-varying protein concentrations
as random variables are reported in \cite{yi2008,steuer2004}. They
provide a reasonable explanation for the size control in wild type
and the quantised CTs in the double-mutant type. Both models employ
the Langevin approach and hence require many simulation runs to provide
an ensemble for computing the mean and (co)variance. However, the
simulation results of stochastic models in \cite{sveiczer2000,sveiczer2002,sveiczer2001,steuer2004,yi2008}
represent one trajectory (for a large number of successive cycles)
of the many possible in the ensemble from which the CT statistics
(time averages) are computed. We will see that the time-averages computed
from the 2MA simulation are for the ensemble of all trajectories.

\subsection{The deterministic model}

We base our 2MA model on the deterministic ODE model for the fission
yeast cell cycle, developed by Tyson-Novák in \cite{novak2001}. In
this context, the cell cycle control mechanism centres around the
M-phase promoting factor (MPF), the active form of the heterodimer
Cdc13/Cdc2, and its antagonistic interactions with enemies (Ste9,Slp1,Rum1)
and the positive feedback with its friend Cdc25. These interactions,
among many others, define a sequence of check points to control the
timing of cell cycle phases. The result is MPF activity oscillation
between low (G1-phase), intermediate (S- and G2-phases) and high (M-phase)
levels that is required for the correct sequence of cell cycle events.
For simplicity, it is assumed that the cell divides functionally when
MPF drops from 0.1.

Table \ref{tab:proteins-rates} lists the proteins whose concentrations
$x_{i}$, together with MPF concentration, are treated as dynamic
variables that evolve according to\begin{equation}
\frac{dx_{i}}{dt}=f_{i}^{+}(x)-f_{i}^{-}(x)\,.\label{eq:ode-xi}\end{equation}
Here $f_{i}^{+}(x)$ is the production flux and \foreignlanguage{english}{$f_{i}^{-}(x)$
is the elimination flux of $i$th protein. Note that the summands
in the fluxes }$f_{i}^{+}(x)$\foreignlanguage{english}{ and $f_{i}^{-}(x)$
are rates of reactions, most of which, are non-elementary (summarizing
many elementary reactions into a single step). Quite a few of these
reaction rates have rational expressions which requires the extended
2MA approach developed in this paper. The MPF concentration $x_{\mathrm{mpf}}$
can be obtained from the algebraic relation\begin{equation}
x_{\mathrm{mpf}}=\frac{\left(x_{1}-x_{2}\right)\left(x_{1}-x_{\mathrm{trim}}\right)}{x_{1}}\label{eq:mpf}\end{equation}
where\begin{equation}
\begin{aligned}\frac{dM}{dt} & =\rho M\\
x_{\mathrm{trim}} & =\frac{2x_{1}x_{7}}{\mathit{\Sigma}+\sqrt{\mathit{\Sigma}^{2}-4x_{1}x_{7}}}\\
x_{\mathrm{tf}} & =G\left(k_{15}M,k'_{16},k''_{16}x_{\mathrm{mpf}},J_{15},J_{16}\right)\\
k_{\mathrm{wee}} & =k'_{\mathrm{wee}}+\left(k''_{\mathrm{wee}}-k'_{\mathrm{wee}}\right)G\left(V_{\mathrm{awee}},V_{\mathrm{iwee}}x_{\mathrm{mpf}},J_{\mathrm{awee}},J_{\mathrm{iwee}}\right)\\
k_{25} & =k'_{25}+\left(k''_{25}-k'_{25}\right)G\left(V_{\mathrm{a}25}x_{\mathrm{mpf}},V_{\mathrm{i}25},J_{\mathrm{a}25},J_{\mathrm{i}25}\right)\\
\mathit{\Sigma} & =x_{1}+x_{7}+K_{\mathrm{diss}},\\
G(a,b,c,d) & =\frac{2ad}{b-a+bc+ad+\sqrt{(b-a+bc+ad)^{2}-4(b-a)ad}}\end{aligned}
\label{eq:aux}\end{equation}
Note that }the cellular mass $M$ is assumed to grow exponentially
with a rate $\rho$, and the concentrations ($x_{\mathrm{trim}},x_{\mathrm{tf}},k_{\mathrm{wee}},k_{25}$)
are assumed to be in a pseudo-steady-state to simplify the model.
Note that we use a slightly different notation: $\rho$ for mass growth
rate (instead of $\mu$), $x_{\mathrm{trim}}$ for Trimmer concentration
and $x_{\mathrm{tf}}$ for TF concentration. We have to emphasise
that the concentrations used in this model are relative and dimensionless.
When one concentration is divided by another, the proportion is the
same as a proportion of two copy numbers. Hence, such a concentration
should not be interpreted as a copy number per unit volume (as misinterpreted
in \cite{yi2008}). \foreignlanguage{english}{The parameters used
in the }Tyson-Novák model \foreignlanguage{english}{\cite{novak2001}}
are\foreignlanguage{english}{ listed in Table~\ref{tab:param} in
Appendix A3.}

\selectlanguage{english}%
\begin{table}
\begin{centering}
\caption{\label{tab:proteins-rates}Proteins and fluxes. Here $x$ denotes
the vector of concentrations $x_{1}$ to $x_{8}$.}
\smallskip{}

\par\end{centering}

\centering{}\begin{tabular}{|c|c|c|c|}
\hline 
 Index & \selectlanguage{british}%
Protein\selectlanguage{english}
 & Production flux & Elimination flux\tabularnewline
\selectlanguage{british}%
$i$\selectlanguage{english}
 & \selectlanguage{british}%
\selectlanguage{english}
 & $f_{i}^{+}(x)$ & $f_{i}^{-}(x)$\tabularnewline
\hline
\hline 
\selectlanguage{british}%
1\selectlanguage{english}
 & Cdc13$_{\text{T}}$ & \selectlanguage{british}%
$k_{1}M$\selectlanguage{english}
 & $\left(k'_{2}+k''_{2}x_{3}+k'''_{2}x_{5}\right)x_{1}$\tabularnewline
\hline 
2 & preMPF & $\left(x_{1}-x_{2}\right)k_{\mathrm{wee}}$ & $\left(k_{25}+k'_{2}+k''_{2}x_{3}+k'''_{2}x_{5}\right)x_{2}$\tabularnewline
\hline 
3 & Ste9 & $\frac{\left(k'_{3}+k''_{3}x_{5}\right)\left(1-x_{3}\right)}{J_{3}+1-x_{3}}$ & $\frac{\left(k'_{4}x_{8}+k_{4}x_{\mathrm{mpf}}\right)x_{3}}{J_{4}+x_{3}}$\tabularnewline
\hline 
4 & Slp1$_{\text{T}}$ & $k'_{5}+\frac{k''_{5}x_{\mathrm{mpf}}^{4}}{J_{4}^{4}+x_{\mathrm{mpf}}^{4}}$ & $k_{6}x_{4}$\tabularnewline
\hline 
5 & Slp1 & $k_{7}\frac{\left(x_{4}-x_{5}\right)x_{6}}{J_{7}+x_{4}-x_{5}}$ & $k_{6}x_{5}+k_{8}\frac{x_{5}}{J_{8}+x_{5}}$\tabularnewline
\hline 
6 & IEP & $k_{9}\frac{\left(1-x_{6}\right)x_{\mathrm{mpf}}}{J_{9}+1-x_{6}}$ & $k_{10}\frac{x_{6}}{J_{10}+x_{6}}$\tabularnewline
\hline 
7 & Rum1$_{\text{T}}$ & $k_{11}$ & $\left(k_{12}+k'_{12}x_{8}+k''_{2}x_{\mathrm{mpf}}\right)x_{7}$\tabularnewline
\hline 
8 & SK & $k_{13}x_{\mathrm{tf}}$ & $k_{14}x_{8}$\tabularnewline
\hline
\end{tabular}
\end{table}

\selectlanguage{british}%
The deterministic ODE model describes the behaviour of an \textquoteleft{}average
cell\textquoteright{}, neglecting the differences among cells in culture.
\foreignlanguage{english}{Specifically, it fails to explain the experimentally
observed clusters of the CT-vs-BM plot and the tri-modal distribution
of CT }\cite{sveiczer1996,sveiczer2000,sveiczer2001,sveiczer2002}\foreignlanguage{english}{.}

\subsection{Feasibility of Gillespie simulations}

Ideally, we should repeat many runs of Gillespie's SSA and compute
our desired moments from the ensemble of those runs. At present, there
are two problems which this. The first problem is the requirement
of elementary reactions for SSA. The elementary reactions underlying
the deterministic model \cite{novak2001} are not known. Many elementary
steps have been simplified to obtain that model. Trying to perform
SSA on non-elementary reactions will lose the discrete event character
of SSA. The second problem arises from the fact that the SSA requires
copy numbers which in turn requires knowledge of measured concentrations.
All protein concentrations in the model are expressed in arbitrary
units (a.u.) because the actual concentrations of most regulatory
proteins in the cell are not known~\cite{csikasz-nagy2006}. Tyson
and Sveiczer%
\footnote{Personal communication.%
} define relative concentration $x_{i}$ of the $i$th protein as $x_{i}=n_{i}/\mathit{\Omega}_{i}$
where $\mathit{\Omega}_{i}=C_{i}N_{A}V$. Here $C_{i}$ is an unknown
characteristic concentration of the $i$th component. The idea is
to make the relative concentrations $x_{i}$ free of scale of the
actual (molar) concentrations $n_{i}/N_{A}V$. Although one would
like to vary $C_{i}$, this is computationally intensive. This problem
is not so serious for the continuous approximations such as CLE, LNA
and the 2MA which are all ODEs and can be numerically solved.

\selectlanguage{english}%

\subsection{The stochastic model using Langevin's approach}

In \cite{yi2008} a stochastic model is proposed that replaces the
ODE model \eqref{eq:ode-xi} with a set of chemical Langevin equations
(CLEs)\foreignlanguage{british}{\[
\frac{d}{dt}x_{i}(t)=f_{i}^{+}\bigl(x(t)\bigr)-f_{i}^{-}\bigl(x(t)\bigr)+\frac{1}{\mathit{\Omega}}\left[\sqrt{f_{i}^{+}\left(x(t)\right)}\mathit{\Gamma}_{i}^{+}(t)-\sqrt{f_{i}^{-}\left(x(t)\right)}\mathit{\Gamma}_{i}^{-}(t)\right],\]
which uses the Langevin noise terms: White noises} $\Gamma_{i}^{+}$
and $\Gamma_{i}^{-}$ scaled by $\sqrt{f_{i}^{+}(x)}$ and $\sqrt{f_{i}^{-}(x)}$
to represent the internal noise. The system parameter $\mathit{\Omega}$
has been described as the volume by the author. As we discussed before,
the concentrations are relative levels with different system size
parameters. That means that concentrations are not the same as copy
numbers per unit volume.

Another stochastic model employing the Langevin's approach is reported
in \cite{steuer2004} which approximates the squared noise amplitudes
by linear functions: \foreignlanguage{british}{\[
\frac{d}{dt}x_{i}(t)=f_{i}\left(x(t)\right)+\sqrt{2D_{i}x_{i}(t)}\mathit{\Gamma}_{i}(t),\]
where $D_{i}$ is a constant. }The reason why the model dynamics $f(x)$
are missing in this model is that the author wanted to represent both
the internal and external noise by the second term on the right.

\subsection{The 2MA model}

\selectlanguage{british}%
For the cell cycle model, the flux $f$ and the diffusion matrix $B$,
defined in \eqref{eq:2MA-aux}, have elements\[
f_{i}(x)=f_{i}^{+}(x)-f_{i}^{-}(x),\quad B_{ik}(x)=\begin{cases}
f_{i}^{+}(x)+f_{i}^{-}(x) & \,\mbox{if}\, i=k\\
0 & \,\mbox{if}\, i\ne k\,.\end{cases}\]
The off-diagonal elements of $B$ are zero because each reaction changes
only one component, so that $S_{ij}S_{kj}=0$ for $i\ne k$. Once
these quantities are known, it follows from \eqref{eq:2MA-mu} and
\eqref{eq:2MA-sigma} that the following set of ODEs: \begin{align}
\frac{d\mu_{i}}{dt} & =f_{i}(\mu)+\varepsilon_{f_{i}}(\mu,\sigma)\label{eq:2MA-mui}\\
\frac{d\sigma_{ii}}{dt} & =2\sum_{l}A_{il}(\mu)\sigma_{li}+\frac{1}{\mathit{\Omega}_{i}}\left[B_{ii}(\mu)+\varepsilon_{B_{ii}}(\mu,\sigma)\right]\label{eq:2MA-sigmaii}\\
\frac{d\sigma_{ik}}{dt} & =\sum_{l}\left[A_{il}(\mu)\sigma_{lk}+\sigma_{il}A_{kl}(\mu)\right]\quad i\ne k\,\label{eq:2MA-sigmaik}\end{align}
approximates (correctly to the 2nd order moments) the evolution of
component-wise concentration mean and covariance. See See Tables \ref{tab:jacobean}-\ref{tab:diffusion-correct}
in Appendix A3 for the respective expressions of the drift matrix
$A$, the stochastic flux $\varepsilon_{f}$ and the correction-term
$\varepsilon_{B}$ added to the diffusion matrix $B$ in \eqref{eq:2MA-sigmaii}.

Having at hand the moments involving the eight dynamic variables $x_{1}$
to $x_{8}$, the mean MPF concentration can be shown to be approximately
(correct to 2nd order moments):

\begin{equation}
\mu_{\mathrm{mpf}}=\mu_{1}-\mu_{2}-x_{\mathrm{trim}}+\frac{x_{\mathrm{trim}}}{\mu_{1}}\left[\left(1+\frac{\sigma_{11}}{\mu_{1}^{2}}\right)\mu_{2}-\frac{\sigma_{12}}{\mu_{1}}\right]\label{eq:mu_mpf}\end{equation}
for the mean MPF concentration with the understanding that $x_{\mathrm{trim}}$
is in pseudo steady state (See Appendix A2 for the derivation). This
expression for the average MPF activity demonstrates the influence
of (co)variance on the mean as emphasised here. We see the dependence
of mean MPF concentration $\mu_{\mathrm{mpf}}$ on the variance $\sigma_{11}$
and covariance $\sigma_{12}$ in addition to the means $\mu_{1},\mu_{2}$
and $x_{\mathrm{trim}}$.

\subsection{Simulations of the 2MA model}

\begin{figure}
\begin{centering}
\hfill{}\subfloat[\label{fig:mean-wt}]{\includegraphics[width=0.48\columnwidth]{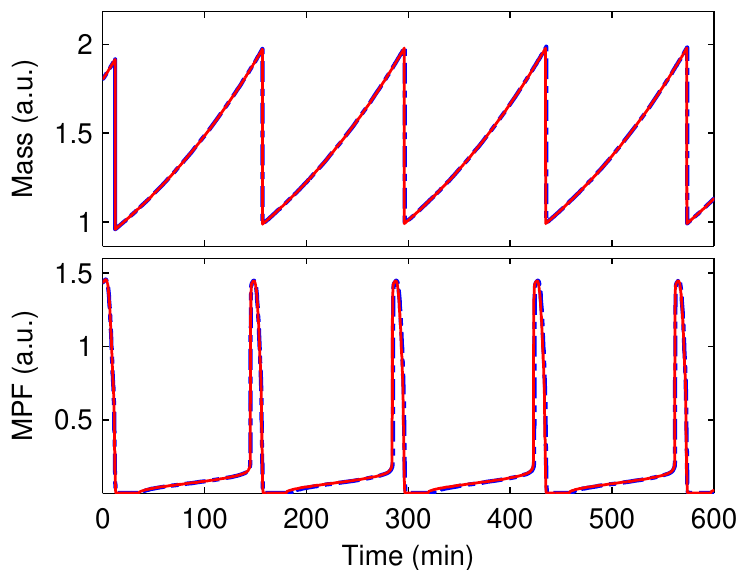}}\hfill{}\subfloat[\label{fig:mean-dm}]{\includegraphics[width=0.48\columnwidth]{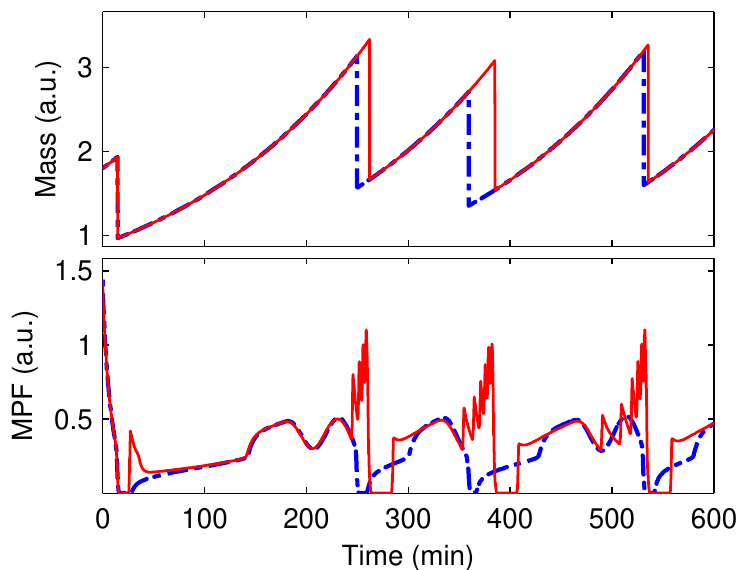}}\hfill{}
\par\end{centering}

\centering{}\caption{\label{fig:mean}The time-courses of mass and MPF activity: (a) for
the wild type, (b) for the double-mutant type. The 2MA predicted mean
trajectories are plotted as solid lines and the corresponding deterministic
trajectories as dashed lines. The difference between the two predictions
is negligible for the wild type, but significant for double-mutant
type.}

\end{figure}

The system of ODEs \eqref{eq:2MA-mui}-\eqref{eq:2MA-sigmaik} was
solved numerically by the MATLAB solver ode15s \cite{themathworks2007}.
The solution was then combined with algebraic relations \eqref{eq:mu_mpf}.
For parameter values, see Table~\ref{tab:param}. Since information
about the individual scaling parameters $\mathit{\Omega}_{i}$ used
in the definition of concentrations is not available, we have used
$\mathit{\Omega}_{i}=5000$ for all $i$. This value has also been
used in \cite{yi2008}, although there is no clear justification.
Note, however, that the 2MA approach developed here will work for
any combination of $\left\{ \mathit{\Omega}_{i}\right\} $. The time-courses
of mass and MPF activity are plotted in Figure~\ref{fig:mean-wt}
for the wild type and in Figure~\ref{fig:mean-dm} for the double-mutant
type. For the wild type, the 2MA predicted mean trajectories do not
differ considerably from the corresponding deterministic trajectories.
Both plots show a more or less constant CT near 150 min. Thus internal
noise does not seem to have a major influence for the wild type.

For the double-mutant type, the difference between the 2MA and deterministic
predictions is significant. The deterministic model \eqref{eq:ode-xi}
predicts alternating short cycles and long cycles because cells born
at the larger size have shorter cycle, and smaller newborns have longer
cycles \cite{novak2001}. This strict alternation due to size control
is not observed in experiments: cells of same mass may have short
or long cycles (excluding very large cells that have always the shortest
CT) \cite{sveiczer1996,sveiczer2000}. This lack of size control is
reproduced by the 2MA simulations: the multiple resettings of MPF
to G2, induced by the internal noise, result in longer CTs (thus accounting
for the 230-min cycles observed experimentally). Such MPF resettings
have been proposed in \cite{sveiczer2000,sveiczer2002} to explain
quantised CTs. No such resetting is demonstrated by the deterministic
model.

Note that the mean $\mu(t)$ of the 2MA describes the average of an
ensemble of cells. Yet the MPF resettings observed in Figure \eqref{fig:mean-dm},
near G2/M transition, introduce the required variability that explains
the clustering of the cycle time observed in experiments. This is
in contrast to the alternative stochastic approaches in \cite{sveiczer2000,sveiczer2002,sveiczer2001,steuer2004,yi2008}
that use one sample trajectory rather than the ensemble average.

How do we explain this significant effect of noise for the double-mutant
on one hand and its negligible effect for the wild type on the other
hand? If we look at expression \eqref{eq:mu_mpf}, we see the influence
of the variance $\sigma_{11}$ (of \foreignlanguage{english}{Cdc13$_{\text{T}}$})
and covariance $\sigma_{12}$ (between \foreignlanguage{english}{Cdc13$_{\text{T}}$}
and preMPF) on the mean MPF concentration $\mu_{\mathrm{mpf}}$. %
\begin{figure}
\begin{centering}
\hfill{}\subfloat[\label{fig:cov-wt}]{\includegraphics[width=0.48\columnwidth]{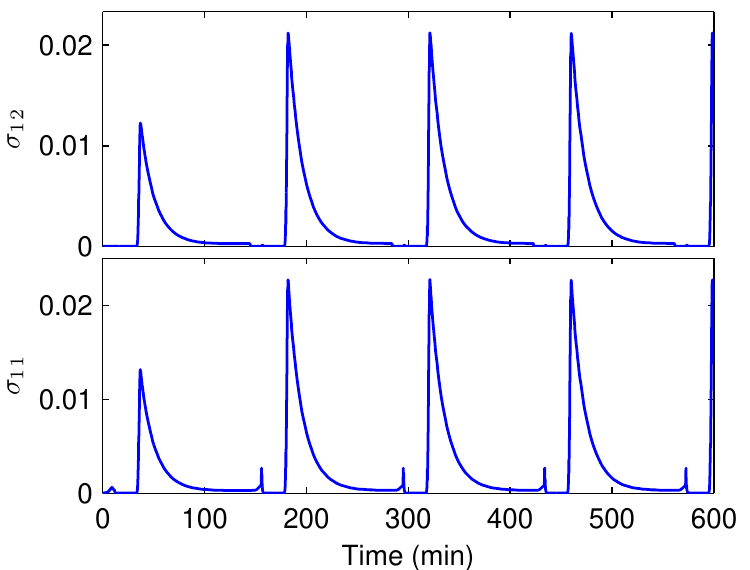}}\hfill{}\subfloat[\label{fig:cov-dm}]{\includegraphics[width=0.48\columnwidth]{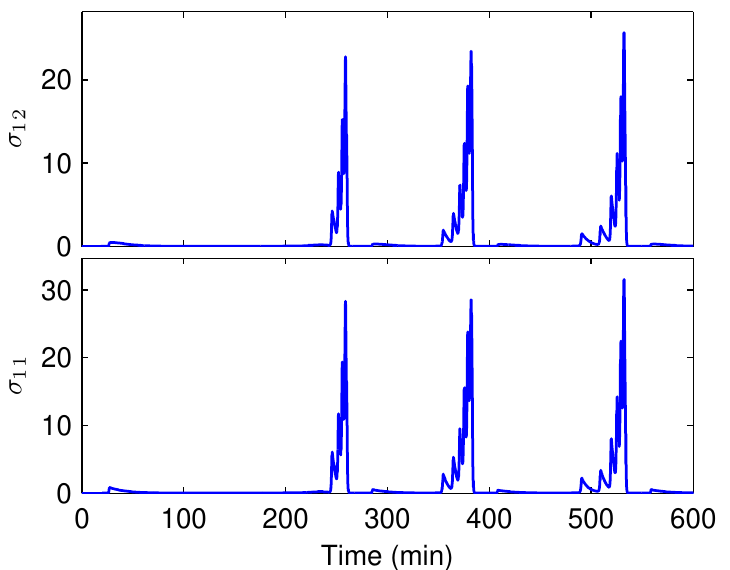}}\hfill{}
\par\end{centering}

\caption{\label{fig:cov}Variance $\sigma_{11}$ (of \foreignlanguage{english}{Cdc13$_{\text{T}}$})
and covariance $\sigma_{12}$ (between \foreignlanguage{english}{Cdc13$_{\text{T}}$}
and preMPF): (a) for the wild type, (b) for double-mutant type.}

\end{figure}
The two (co)variances are plotted in Figure~\ref{fig:cov-wt} for
the wild type and in Figure~\ref{fig:cov-dm} for the double-mutant
type. It is clear that the two (co)variances have very small peaks
for the wild type compared to the large peaks for the double-mutant
type. Note that the larger peaks in Figure~\ref{fig:cov-dm} are
located at the same time points where the MPF activity exhibits oscillations
and hence multiple resettings to G2. This suggest that the oscillatory
behaviour of MPF near the G2/M transition is due to the influence
of the oscillatory (co)variances. This coupling between the mean and
(co)variance is not captured by the deterministic model.

\begin{figure}
\begin{centering}
\hfill{}\subfloat[]{\includegraphics[width=0.48\columnwidth]{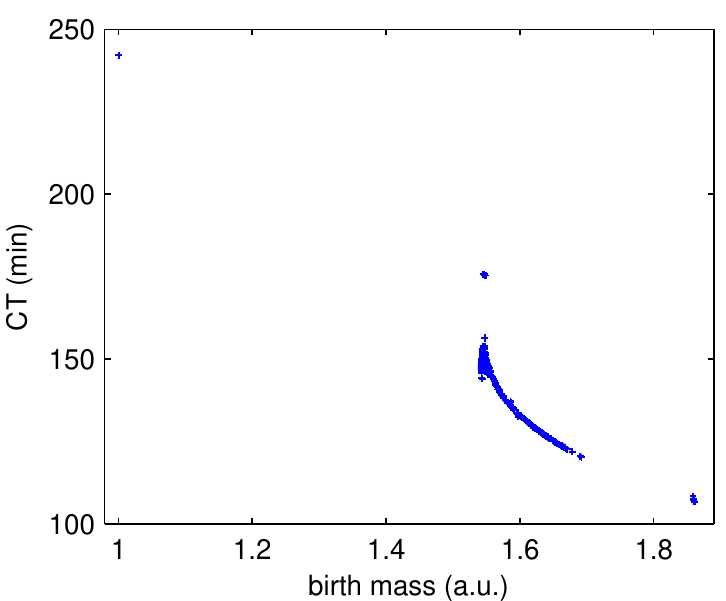}}\hfill{}\subfloat[]{\includegraphics[width=0.48\columnwidth]{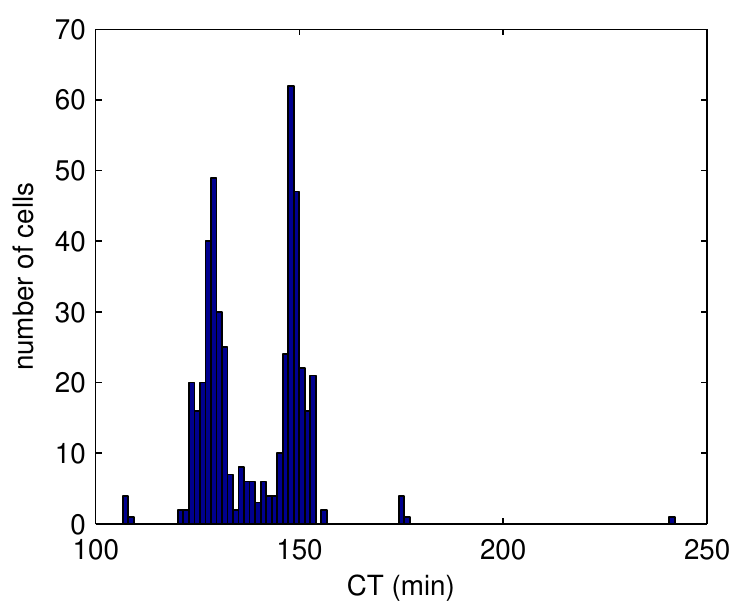}}\hfill{}
\par\end{centering}

\begin{centering}
\hfill{}\subfloat[]{\includegraphics[width=0.48\columnwidth]{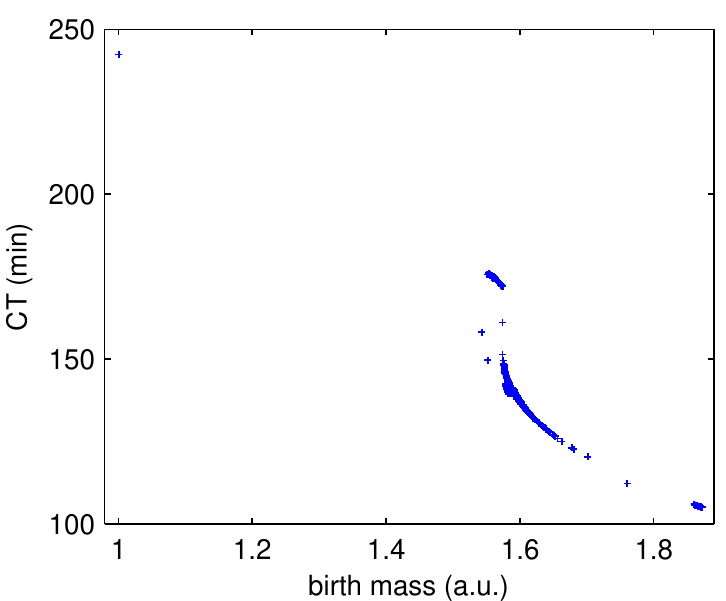}}\hfill{}\subfloat[]{\includegraphics[width=0.48\columnwidth]{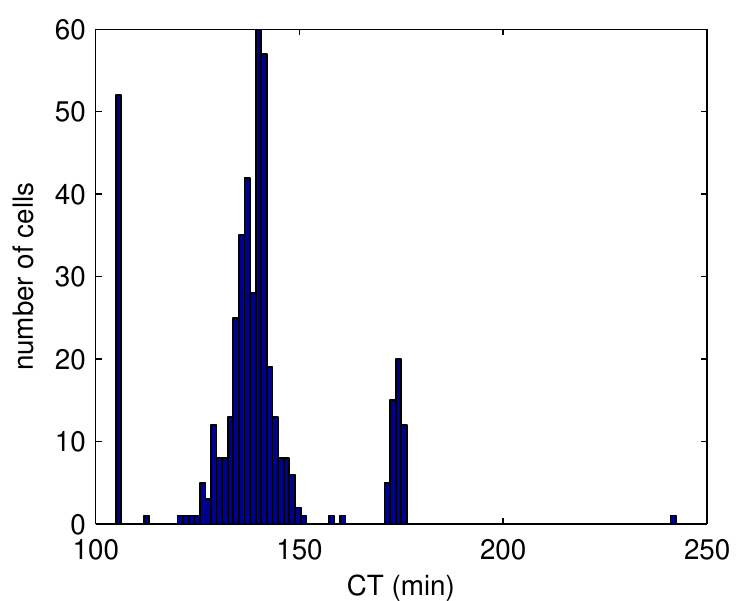}}\hfill{}
\par\end{centering}

\begin{centering}
\hfill{}\subfloat[]{\includegraphics[width=0.48\columnwidth]{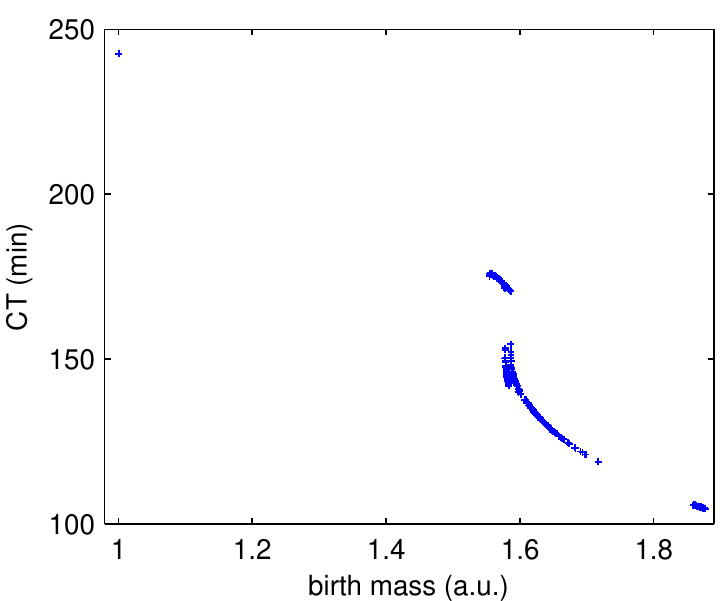}}\hfill{}\subfloat[]{\includegraphics[width=0.48\columnwidth]{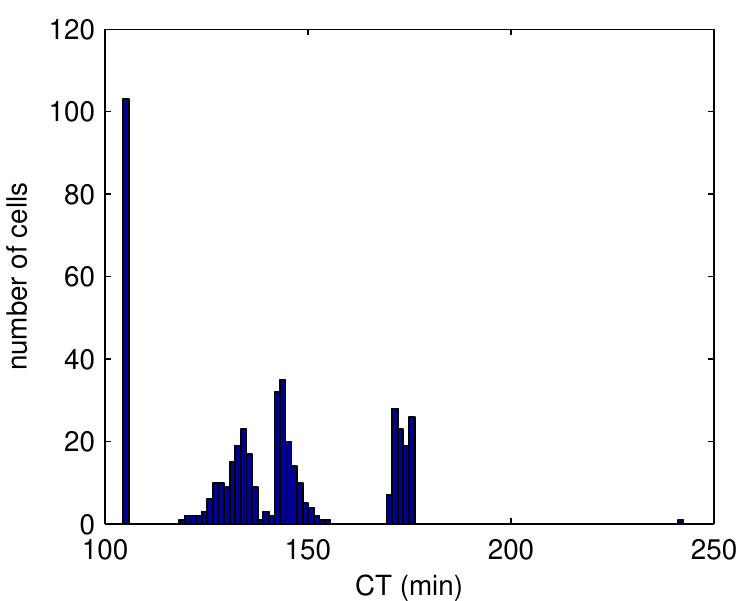}}\hfill{}
\par\end{centering}

\caption{\label{fig:CT-stats}Cycle time behaviour over 465 successive cycles
of the double-mutant cells, predicted by the 2MA model.\emph{ }(a,c,e):
CT vs BM, (b,d,f): CT distribution, (a,b): $\mathit{\Omega}=5000$,
(c,d): $\mathit{\Omega}=5200$, (e,f): $\mathit{\Omega}=5300$. The
plots are in qualitative agreement to experiments, see \cite[Figure 6]{sveiczer1996}
and \cite[Figure 5]{sveiczer2002} for a comparison.}

\end{figure}

It has to be realised that the above proposition requires validation
since the 2MA approach ignores 3rd and higher-order moments. We cannot
know whether that truncation is responsible for the oscillations in
Figures \ref{fig:mean} and \ref{fig:cov}, unless compared with a
few sample trajectories simulated by the SSA. However, as discussed
before, the SSA cannot be performed (at present) for the model in
consideration. Therefore we need to compare the 2MA predictions for
the double-mutant type cells with experimental data. Towards that
end, values of cycle time (CT), birth mass (BM) and division mass
(DM) were computed for 465 successive cycles of double-mutant cells.
Figure~\ref{fig:CT-stats} shows the CT-vs-BM plot and the CT distribution
for three different values $\{5000,5200,5300\}$ of system size $\mathit{\Omega}$.

 To make this figure comparable with experimental data from \cite{sveiczer1996,sveiczer2002},
we assume that 1 unit of mass corresponds to 8.2 $\mu$m cell length
\cite{sveiczer2000}. We can see the missing size control (CT clusters),
in qualitative agreement with experimentally observed ones (see \cite[Figure 6]{sveiczer1996}
and \cite[Figure 5]{sveiczer2002} for a comparison). There are more
than four clusters, which may have arisen from the truncated higher-order
moments. The extreme value of CT higher than 230 min suggests more
than two MPF resettings. Furthermore, more than three modes in the
CT distribution may have arisen from the truncated higher-order moments.
\begin{table}
\begin{centering}
\caption{\label{tab:stats}Statistics over 465 successive cell cycles of the
double-mutant type cells, predicted by the 2MA model, compared with
experimental data, see \cite[Table 1]{sveiczer1996}.}
\smallskip{}

\par\end{centering}

\centering{}\begin{tabular}{|c|c|c|c|c|c|c|c|c|}
\hline 
Case & $\mu_{\mathrm{CT}}$ & $\sigma_{\mathrm{CT}}$ & $\mathrm{CV_{CT}}$ & $\mu_{\mathrm{DM}}$ & $\sigma_{\mathrm{DM}}$ & $\mathrm{CV_{DM}}$ & $\mu_{\mathrm{BM}}$ & $\sigma_{\mathrm{BM}}$\tabularnewline
\hline
\hline 
(1) & 131 & 47 & 0.358 & 2.22 & 0.45 & 0.203 & 1.21 & 0.24\tabularnewline
\hline 
(2) & 138.8 & 12.4 & 0.09 & 3.18 & 0.101 & 0.0319 & 1.59 & 0.0575\tabularnewline
\hline 
(3) & 138.8 & 17.6 & 0.127 & 3.25 & 0.178 & 0.055 & 1.623 & 0.0934\tabularnewline
\hline 
(4) & 138.8 & 23.9 & 0.172 & 3.32 & 0.231 & 0.0697 & 1.657 & 0.12\tabularnewline
\hline
\end{tabular}\smallskip{}
\\
(1) experimental data, (2) $\mathit{\Omega}=5000$, (3) $\mathit{\Omega}=5200$,
(4) $\mathit{\Omega}=5300$.
\end{table}
Table \ref{tab:stats} compares the statistics for the double-mutant
type cells, computed with the 2MA approach, with data from \cite[Table 1]{sveiczer1996}.
Column 2-4 tabulate, for CT, the mean $\mu_{\mathrm{CT}}$, the standard
deviation $\sigma_{\mathrm{CT}}$ and the coefficient of variation
$\mathrm{CV_{CT}}$, respectively. The other columns tabulate similar
quantities for the division mass (DM) and the birth mass (BM). We
see that only the mean CT is in agreement with the experimental data.
The mean values for both BM and DM are larger than the corresponding
experimental values. The other statistics are much smaller the corresponding
experimental values. This and the above plots suggest that the 2MA
should be used with caution. However, another aspect of the cell cycle
model deserves attention here. The way the relative protein concentrations
have been defined implies unknown values of the scaling parameters
$\{\mathit{\Omega}_{i}\}$. Since $\mathit{\Omega}_{i}=C_{i}N_{A}V$,
knowing the volume $V$ does not solve the problem: the characteristic
concentrations $\{C_{i}\}$ are still unknown. Our simulations have
chosen typical values $\mathit{\Omega}=\{5000,5200,5300\}$. The corresponding
three pairs of plots in Figure~\ref{fig:CT-stats} and rows in Table~\ref{tab:stats}
demonstrate a dependence of the results on a suitable system size.
There is no way to confirm these values. The scaling parameters could
be regulated in a wider range in order to imporve the accuracy of
our simulation, motivating future work for us. The conclusion is that
the quantitative disagreement of the 2MA predictions can be attributed
to two factors: 1) the truncated higher-order moments during the derivation
of the 2MA, and (2) the unknown values of scaling parameters.

\section{Conclusions}

The recently developed two-moment approximation (2MA) \cite{goutsias2007,gomez-uribe2007}
is a promising approach because it accounts for the coupling between
the means and (co)variances. We have extended the derivation of the
2MA to biochemical networks and established two advances to previous
efforts: a) relative concentrations and b) non-elementary reactions.
Both aspects are important in systems biology where one is often forced
to aggregate elementary reactions into single step reactions. In these
situations one cannot assume knowledge of elementary reactions to
formulate a stochastic model. Previous derivations assumed elementary
reactions and absolute concentrations. However, numerous existing
models in systems biology use relative concentrations. 

We investigated the applicability of the 2MA approach to the well
established fission yeast cell cycle model. The simulations of the
2MA model show oscillatory behaviour near the G2/M transition, which
is significantly different from the simulations of deterministic ODE
model. One notable aspect of our analytical model is that, although
it describes the average of an ensemble, it reproduces enough variability
among cycles to reproduce the curious quantised cycle times observed
in experiments on double mutants.
\begin{ack}
In the process of preparing this manuscript Hanspeter Herzel (Humboldt
University), Akos Sveiczer (Budapest University of Technology and
Economics) and Kevin Burrage (University of Queensland, Brisbane)
helped in clarifying questions related to the cell cycle and stochastic
modelling. We very appreciate their willingness to have discussed
these issues with thus. M.U. has been supported by the Deutsche Forschungsgemeinschaft
(DFG) through grant (WO 991/3-1). O.W. acknowledges support by the
Helmholtz Alliance on Systems Biology and the Stellenbosch Institute
for Advanced Study (STIAS).
\end{ack}
\selectlanguage{english}%
\appendix
\appendixpage
\renewcommand{\thesection}{A\arabic{section}}
\setcounter{section}{0}

\selectlanguage{british}%

\section{Derivation of the 2MA equations}

The progress of a particular reaction can be described by a quantity
known as the \emph{degree of advancement} (DA). We will write $Z_{j}(t)$
for the DA of the $j$th reaction, where $Z_{j}(t)=z_{j}$ means that
the $j$th reaction has occurred $z_{j}$ times during the interval
$[0,t)$. In the same interval the $j$th reaction will contribute
a change of $z_{j}S_{ij}$ molecules to the overall change in the
copy number $N_{i}$ of the $i$th component. Summing up contributions
from all the reactions, the copy number can be expressed as\begin{equation}
N_{i}(t)=N_{i}(0)+\sum_{j=1}^{r}S_{ij}Z_{j}(t)\,.\label{eq:Ni}\end{equation}
Based on the definition of reaction propensity, the number of occurrences
$Z_{j}(t+\Delta t)-Z_{j}(t)$ during a short interval $[t,t+\Delta t]$
has the probability distribution \begin{multline}
\Pr\left[Z_{j}(t+\Delta t)-Z_{j}(t)=z_{j}\,|\, N(t)=n\right]\\
=\begin{cases}
a_{j}(n)\Delta t+o(\Delta t) & \quad\mbox{if}\quad z_{j}=1\\
1-a_{j}(n)\Delta t+o(\Delta t) & \quad\mbox{if}\quad z_{j}=0\\
o(\Delta t) & \quad\mbox{if}\quad z_{j}>1\end{cases}\label{eq:Pr-del-Zj}\end{multline}
where $o(\Delta t)$ represents a quantity that vanishes faster than
$\Delta t$ as the later approaches zero. In effect, \eqref{eq:Pr-del-Zj}
gives the conditional probability distribution, in state $n$, of
the random progress (DA increment) $Z_{j}(t+\Delta t)-Z_{j}(t)$ of
the $j$th reaction during the time interval $[t,t+\Delta t)$. The
expected value of this short-time DA increment can be obtained from
\eqref{eq:Pr-del-Zj} as \begin{multline}
\mathrm{E}\left[Z_{j}(t+\Delta t)-Z_{j}(t)\,|\, N(t)=n\right]\\
\begin{split} & =\sum_{j=0}^{r}z_{j}\Pr\left[Z_{j}(t+\Delta t)-Z_{j}(t)=z_{j}\,|\, N(t)=n\right]\\
 & =\overbrace{a_{j}(n)\Delta t}^{z_{j}=1}+\overbrace{o(\Delta t)}^{z_{j}>1}\end{split}
\label{eq:EZj}\end{multline}
which is conditioned on $N(t)=n$. The unconditional expectation of
the DA increment can be obtained by summing the probabilities $P(n,t)$
weighted by the above conditional expectation over all possible states
$n$:\[
\begin{split}\mathrm{E}\left[Z_{j}(t+\Delta t)-Z_{j}(t)\right] & =\sum_{n}\mathrm{E}\left[Z_{j}(t+\Delta t)-Z_{j}(t)\,|\, N(t)=n\right]P(n,t)\\
 & =\sum_{n}a_{j}(n)p_{n}(t)\Delta t+o(\Delta t)\\
 & =\mathrm{E}\left[a_{j}\bigl(N(t)\bigr)\right]\Delta t+o(\Delta t)\end{split}
\]
which for vanishingly small $\Delta t$ leads to the ODE\begin{equation}
\frac{d}{dt}\mathrm{E}\left[Z_{j}(t)\right]=\mathrm{E}\left[a_{j}\left(N(t)\right)\right]\label{eq:ode-z_j}\end{equation}
Thus the mean propensity of a particular reaction can be interpreted
as the \emph{average number of occurrences (DA) per unit time} of
that reaction. Take the expectation on both side of the conservation
\eqref{eq:Ni} to obtain \[
\frac{d}{dt}\mathrm{E}\left[N_{i}(t)\right]=\sum_{j=1}^{r}S_{ij}\mathrm{E}\left[a_{j}\bigl(N(t)\bigr)\right]\]
which proves \eqref{eq:mean-copynumber} in the main text. It is interesting
to note that the above ODE is a direct consequence of mass conservation
\eqref{eq:Ni} and definition of propensity because we have not referred
to the CME (which is the usual procedure) during our derivation.

Dividing \eqref{eq:mean-copynumber} by $\mathit{\Omega}_{i}$ gives
the ODE for the component mean concentration,\begin{equation}
\frac{d}{dt}\mu_{i}(t)=\mathrm{E}\left[f_{i}\bigl(X(t)\bigr)\right]\label{eq:mudot_i}\end{equation}
where \[
f_{i}(x)=\frac{1}{\mathit{\Omega}_{i}}\sum_{j=1}^{r}S_{ij}a_{j}(\mathit{\Omega}x)\]
is the total \emph{flux }of component $i$ in state $x$.

Suppose the propensity $a_{j}(n)$ is a smooth function and that central
moments $\mathrm{E}\left[(N-\mu)^{m}\right]$ of order higher than
$m=2$ can be ignored. In that case, the Taylor series expansion of
flux $f_{i}(x)$ around the mean is\[
f_{i}(x)=f_{i}(\mu)+\left[\frac{\partial f_{i}}{\partial x^{T}}\right]_{x=\mu}(x-\mu)+\frac{1}{2}(x-\mu)^{T}\left[\frac{\partial^{2}f_{i}}{\partial x\partial x^{T}}\right]_{x=\mu}(x-\mu)+\cdots\,.\]
 Expectation of the 2nd term on the right is zero. Expectation of
the 3rd term can be written as\[
\varepsilon_{f_{i}}(\mu,\sigma)=\frac{1}{2}\sum_{k,l}\left[\frac{\partial^{2}f_{i}}{\partial x_{k}\partial x_{l}}\right]_{x=\mu}\sigma_{kl}\,.\]
Note that the Taylor expansion in powers of $x-\mu$ is more convincing
than that in powers of $n-\mathrm{E}(n)$ because higher-order terms
vanish quicker in the former. Having arrived at this point, ignoring
terms (moments) higher than 2nd order, we can write:\begin{equation}
\frac{d\mu_{i}}{dt}=f_{i}(\mu)+\varepsilon_{f_{i}}(\mu,\sigma)\label{eq:ode-mui}\end{equation}
for mean component concentration and\[
\frac{d\mu}{dt}=f(\mu)+\varepsilon_{f}(\mu,\sigma)\]
for the mean concentration vector. This last equation proves \eqref{eq:2MA-mu}
in the main text. Here the term $\varepsilon_{f}(\mu,\sigma)$ is
the internal noise that arises from the discrete and random nature
of chemical reactions. Note that this term has been derived from the
CME instead of being assumed like external noise. This shows that
knowledge of fluctuations (even if small) is important for a correct
description of the mean. This also indicates an advantage of the stochastic
framework over it deterministic counterpart: starting from the same
assumptions and approximations, the stochastic framework allows us
to see the influence of fluctuation on the mean. Note that the above
equation is exact for systems where no reaction has an order higher
than two because then 3rd and higher derivatives of propensity are
zero.

Before we can see how the covariance $\sigma$ evolves in time, let
us multiply the CME with $n_{i}n_{k}$ and sum over all $n$,\[
\begin{split}\sum_{n}n_{i}n_{k}\frac{dP(n,t)}{dt} & =\sum_{n}n_{i}n_{k}\sum_{j=1}^{r}\left[a_{j}(n-S_{\centerdot j})P(n-S_{\centerdot j},t)-a_{j}(n)P(n,t)\right]\\
 & =\sum_{n}\sum_{j=1}^{r}\biggl[\left(n_{i}+S_{ij}\right)\left(n_{k}+S_{kj}\right)a_{j}(n)P(n,t)-n_{i}n_{k}a_{j}(n)P(n,t)\biggr]\\
 & =\sum_{n}\sum_{j=1}^{r}\left(n_{k}S_{ij}+n_{i}S_{kj}+S_{ij}S_{kj}\right)a_{j}(n)P(n,t)\end{split}
\]
 where dependence on time is implicit for all variables except $n$
and $s$. Dividing by $\mathit{\Omega}_{i}\mathit{\Omega}_{k}$ and
recognising sums of probabilities as expectations,\[
\frac{d\mathrm{E}\left[X_{i}X_{k}\right]}{dt}=\mathrm{E}\left[X_{k}f_{i}(X)\right]+\mathrm{E}\left[X_{i}f_{k}(X)\right]+\frac{1}{\sqrt{\mathit{\Omega}_{i}\mathit{\Omega}_{k}}}\mathrm{E}\left[B_{ik}(X)\right]\]
where $B(x)$ is the \emph{diffusion matrix} with elements \[
B_{ik}(x)=\frac{1}{\sqrt{\mathit{\Omega}_{i}\mathit{\Omega}_{k}}}\sum_{j=1}^{r}S_{ij}S_{kj}a_{j}(\mathit{\Omega}x)\,.\]
 The relation $\sigma_{ik}=\mathrm{E}\left[X_{i}X_{k}\right]-\mu_{i}\mu_{k}$
can be utilised to yield\begin{equation}
\frac{d\sigma_{ik}}{dt}=\mathrm{E}\left[\left(X_{k}-\mu_{k}\right)f_{i}(X)\right]+\mathrm{E}\left[\left(X_{i}-\mu_{i}\right)f_{k}(X)\right]+\frac{1}{\sqrt{\mathit{\Omega}_{i}\mathit{\Omega}_{k}}}\mathrm{E}\left[B_{ik}(X)\right]\label{eq:sigmadot_ik}\end{equation}
 for the covariances between concentrations of component pairs. The
argument of the first expectation in \eqref{eq:sigmadot_ik} has Taylor
expansion\[
f_{i}(x)\left(x_{k}-\mu_{k}\right)=f_{i}(\mu)\left(x_{k}-\mu_{k}\right)+\left[\frac{\partial f_{i}}{\partial x^{T}}\right]_{x=\mu}(x-\mu)\left(x_{k}-\mu_{k}\right)+\cdots\,.\]
Expectation of the first term on the right is zero. Ignoring 3rd and
higher-order moments, the first expectation in \eqref{eq:sigmadot_ik}
is then\[
\mathrm{E}\left[\left(X_{k}-\mu_{k}\right)f_{i}(X)\right]=\sum_{l}A_{il}(\mu)\sigma_{lk}\]
where $A(x)$ is the drift matrix (the Jacobian of $f(x)$) with elements\[
A_{ik}(x)=\frac{\partial f_{i}(x)}{\partial x_{k}}\,.\]
By a similar procedure, the second expectation \eqref{eq:sigmadot_ik}
is\[
\mathrm{E}\left[\left(X_{i}-\mu_{i}\right)f_{k}(X)\right]=\sum_{l}\sigma_{il}A_{il}(\mu),\]
correct to 2nd-order moments. The element $B_{ik}(x)$ of the diffusion
matrix has Taylor expansion\[
B_{ik}(x)=B_{ik}(\mu)+\left[\frac{\partial B_{ik}}{\partial x^{T}}\right]_{x=\mu}(x-\mu)+\frac{1}{2}(x-\mu)^{T}\left[\frac{\partial^{2}B_{ik}}{\partial x\partial x^{T}}\right]_{x=\mu}(x-\mu)+\cdots\,.\]
Taking term-wise expectation, and ignoring 3rd and higher-order moments,
\[
\mathrm{E}\left[B_{ik}(X)\right]=B_{ik}(\mu)+\varepsilon_{B_{ik}}(\mu,\sigma)\]
where\[
\varepsilon_{B_{ik}}(\mu,\sigma)=\frac{1}{2}\sum_{i',k'}\left[\frac{\partial^{2}B_{ik}}{\partial x_{i'}\partial x_{k'}}\right]_{x=\mu}\sigma_{i'k'}\,.\]
Having these results at hand, we can now write\[
\frac{d\sigma_{ik}}{dt}=\sum_{l}\left[A_{il}(\mu)\sigma_{lk}+\sigma_{il}A_{kl}(\mu)\right]+\frac{1}{\sqrt{\mathit{\Omega}_{i}\mathit{\Omega}_{k}}}\left[B_{ik}(\mu)+\varepsilon_{B_{ik}}(\mu,\sigma)\right]\]
for the component-wise covariances. In matrix notation \[
\frac{d\sigma}{dt}=A(\mu)\sigma+\sigma A(\mu)^{T}+\mathit{\Omega}^{-\nicefrac{1}{2}}\left[B(\mu)+\varepsilon_{B}(\mu,\sigma)\right]\left(\mathit{\Omega}^{-\nicefrac{1}{2}}\right)^{T}\]
 proves \eqref{eq:2MA-sigma} in the main text. The \emph{drift matrix}
$A(\mu)$ reflects the dynamics for relaxation (dissipation) to the
steady state and the \emph{diffusion matrix} $B(\mu)$ the randomness
(fluctuation) of the individual events \cite{paulsson2006}. These
terms are borrowed from the \emph{fluctuation-dissipation theorem}
(FDT) \cite{keizer1987,lax1960}, which has the same form as \eqref{eq:2MA-sigma}.
Remember that \eqref{eq:2MA-sigma} is exact for systems that contain
only zero and first-order reactions because in that case the propensity
is already linear.

\section{Mean MPF concentration}

To find the mean MPF concentration, we start with the MPF concentration
\[
x_{\mathrm{mpf}}=\left(x_{1}-x_{2}\right)\left(1-\frac{x_{\mathrm{trim}}}{x_{1}}\right)=x_{1}-x_{2}-x_{\mathrm{trim}}+x_{\mathrm{trim}}\frac{x_{2}}{x_{1}}\,.\]
The ratio $\nicefrac{x_{2}}{x_{1}}$ can be expanded around the mean,
\[
\frac{x_{2}}{x_{1}}=\frac{1}{\mu_{1}}\frac{x_{2}}{1+\frac{\left(x_{1}-\mu_{1}\right)}{\mu_{1}}}=\frac{1}{\mu_{1}}\left[x_{2}-\frac{\left(x_{1}-\mu_{1}\right)x_{2}}{\mu_{1}}+\frac{\left(x_{1}-\mu_{1}\right)^{2}x_{2}}{\mu_{1}^{2}}+\cdots\right]\,.\]
Taking expectation on both sides,\[
\begin{split}\mathrm{E}\left[\frac{X_{2}}{X_{1}}\right] & =\frac{1}{\mu_{1}}\mathrm{E}\left[\frac{X_{2}}{1+\frac{\left(X_{1}-\mu_{1}\right)}{\mu_{1}}}\right]\\
 & =\frac{1}{\mu_{1}}\mathrm{E}\left[X_{2}-\frac{\left(X_{1}-\mu_{1}\right)X_{2}}{\mu_{1}}+\frac{\left(X_{1}-\mu_{1}\right)^{2}X_{2}}{\mu_{1}^{2}}+\cdots\right]\\
 & =\frac{1}{\mu_{1}}\left[\mu_{2}-\frac{\sigma_{12}}{\mu_{1}}+\frac{\mu_{2}\sigma_{11}}{\mu_{1}^{2}}\right]\,.\end{split}
\]
Finally, the mean MPF concentration follows from the expectation of
$x_{\mathrm{mpf}}$ to be \[
\mu_{\mathrm{mpf}}=\mu_{1}-\mu_{2}-x_{\mathrm{trim}}+\frac{x_{\mathrm{trim}}}{\mu_{1}}\left[\left(1+\frac{\sigma_{11}}{\mu_{1}^{2}}\right)\mu_{2}-\frac{\sigma_{12}}{\mu_{1}}\right],\]
thus proving \eqref{eq:mu_mpf} in the main text.

\selectlanguage{english}%

\section{Parameters and coefficients of the 2MA equations}

\begin{table}
\caption{\label{tab:param}Parameter values for the Tyson-Novák cell cycle
model of the fission yeast (wild type) \cite{novak2001}. All constants
have units $\unit{min^{-1}}$, except the $J$s, which are dimensionless
Michaelis constants, and $K_{\mathrm{diss}}$, which is a dimensionless
equilibrium constant for trimer dissociation. For the double-mutant
type, one makes the following three changes: $k''_{\mathrm{wee}}=0.3,\, k'_{25}=k''_{25}=0.02$~.}
\smallskip{}
\framebox{\begin{minipage}[t]{1\columnwidth}%
\begin{flalign*}
 & k_{15}=0.03,\, k'_{2}=0.03,\, k''_{2}=1,\, k'''_{2}=0.1,k'_{3}=1,\, k''_{3}=10,\, J_{3}=0.01,\\
 & k'_{4}=2,\, k_{4}=35,\, J_{4}=0.01,k'_{5}=0.005,\, k''_{5}=0.3,\, k_{6}=0.1,\, J_{5}=0.3,\\
 & k_{7}=1,\, k_{8}=0.25,\, J_{7}=J_{8}=0.001,\, J_{8}=0.001,\, k_{9}=0.1,\, k_{10}=0.04,\\
 & J_{9}=0.01,\, J_{10}=0.01,\, k_{11}=0.1,\, k_{12}=0.01,\, k'_{12}=1,\, k''_{12}=3,\, K_{\mathrm{diss}}=0.001,\\
 & k_{13}=0.1,\, k_{14}=0.1,\, k_{15}=1.5,\, k'_{16}=1,\, k''_{16}=2,\, J_{15}=0.01,\, J_{16}=0.01,\\
 & V_{\mathrm{awee}}=0.25,\, V_{\mathrm{iwee}}=1,\, J_{\mathrm{awee}}=0.01,\, J_{\mathrm{iwee}}=0.01,\, V_{\mathrm{a25}}=1,V_{\mathrm{i25}}=0.25,\\
 & J_{\mathrm{a25}}=0.01,\, J_{\mathrm{i25}}=0.01,\, k'_{\mathrm{wee}}=0.15,\, k''_{\mathrm{wee}}=1.3,\, k'_{25}=0.05,\, k''_{25}=5,\,\rho=0.005.\end{flalign*}
\end{minipage}}
\end{table}
\begin{table}
\begin{centering}
\caption{\label{tab:jacobean}Rows of the drift matrix \foreignlanguage{british}{$A$
of the 2MA cell cycle model. We here use} $\mathrm{e}_{i}$ to denote
the $i$th row of $8\times8$ identity matrix.}
\smallskip{}

\par\end{centering}

\begin{tabular}{|c|c|}
\hline 
Index $i$ & $A_{i\centerdot}(x)=\frac{\partial f_{i}}{\partial x^{T}}$\tabularnewline
\hline
\hline 
1 & $-\left(k'_{2}+k''_{2}x_{3}+k'''_{2}x_{5}\right)\mathrm{e}_{1}-k''_{2}x_{1}\mathrm{e}_{3}-k'''_{2}x_{1}\mathrm{e}_{5}$\tabularnewline
\hline 
2 & $k_{\mathrm{wee}}\mathrm{e}_{1}-\left(k_{\mathrm{wee}}+k_{25}+k'_{2}+k''_{2}x_{3}+k'''_{2}x_{5}\right)\mathrm{e}_{2}-k''_{2}x_{2}\mathrm{e}_{3}-k'''_{2}x_{2}\mathrm{e}_{5}$\tabularnewline
\hline 
3 & \selectlanguage{british}%
-$\left[\frac{\left(k'_{4}x_{8}+k_{4}x_{\mathrm{mpf}}\right)J_{4}}{\left(J_{4}+x_{3}\right)^{2}}+\frac{\left(k'_{3}+k''_{3}x_{5}\right)J_{3}}{\left(J_{3}+1-x_{3}\right)^{2}}\right]\mathrm{e}_{3}+\frac{\left(1-x_{3}\right)k''_{3}}{J_{3}+1-x_{3}}\mathrm{e}_{5}-\frac{k'_{4}x_{3}}{J_{4}+x_{3}}\mathrm{e}_{8}$\selectlanguage{english}
\tabularnewline
\hline 
4 & $-k_{6}\mathrm{e}_{4}$\tabularnewline
\hline 
5 & $\frac{k_{7}J_{7}x_{6}}{\left(J_{7}+x_{4}-x_{5}\right)^{2}}\mathrm{e}_{4}-\left[k_{6}+\frac{k_{7}J_{7}x_{6}}{\left(J_{7}+x_{4}-x_{5}\right)^{2}}+\frac{k_{8}J_{8}}{\left(J_{8}+x_{5}\right)^{2}}\right]\mathrm{e}_{5}+\frac{\left(x_{4}-x_{5}\right)k_{7}}{J_{7}+x_{4}-x_{5}}\mathrm{e}_{6}$\tabularnewline
\hline 
6 & $-\left[\frac{k_{9}x_{\mathrm{mpf}}J_{9}}{\left(J_{9}+1-x_{6}\right)^{2}}+\frac{k_{10}J_{10}}{\left(J_{10}+x_{6}\right)^{2}}\right]\mathrm{e}_{6}$\tabularnewline
\hline 
7 & $-\left(k_{12}+k'_{12}x_{8}+k''_{2}x_{\mathrm{mpf}}\right)\mathrm{e}_{7}-k'_{12}x_{7}\mathrm{e}_{8}$\tabularnewline
\hline 
8 & \selectlanguage{british}%
$-k_{14}\mathrm{e}_{8}$\selectlanguage{english}
\tabularnewline
\hline
\end{tabular}
\end{table}
\begin{table}
\begin{centering}
\caption{\label{tab:flux-correct}Stochastic flux, the correction-term added
to the deterministic flux in \eqref{eq:2MA-mu}.}
\smallskip{}

\par\end{centering}

\begin{tabular}{|c|c|}
\hline 
Index $i$ & $\varepsilon_{f}(x,\sigma)=\frac{1}{2}\sum_{k,l}\frac{\partial^{2}f_{i}}{\partial x_{k}\partial x_{l}}\sigma_{kl}$\tabularnewline
\hline
\hline 
1 & \selectlanguage{british}%
$-k''_{2}\sigma_{13}-k'''_{2}\sigma_{15}$\selectlanguage{english}
\tabularnewline
\hline 
2 & $-k''_{2}\sigma_{23}-k'''_{2}\sigma_{25}$\tabularnewline
\hline 
3 & $\left[\frac{\left(k'_{4}x_{8}+k_{4}x_{\mathrm{mpf}}\right)J_{4}}{\left(J_{4}+x_{3}\right)^{3}}-\frac{\left(k'_{3}+k''_{3}x_{5}\right)J_{3}}{\left(J_{3}+1-x_{3}\right)^{3}}\right]\sigma_{33}-\frac{k''_{3}J_{3}\sigma_{35}}{\left(J_{3}+1-x_{3}\right)^{2}}-\frac{k'_{4}J_{4}\sigma_{38}}{\left(J_{4}+x_{3}\right)^{2}}$\tabularnewline
\hline 
4 & $0$\tabularnewline
\hline 
5 & $\frac{k_{7}J_{7}x_{6}\left(2\sigma_{45}-\sigma_{44}-\sigma_{55}\right)}{\left(J_{7}+x_{4}-x_{5}\right)^{3}}+\frac{k_{7}J_{7}\left(\sigma_{46}-\sigma_{56}\right)}{\left(J_{7}+x_{4}-x_{5}\right)^{2}}+\frac{k_{8}J_{8}}{\left(J_{8}+x_{5}\right)^{3}}\sigma_{55}$\tabularnewline
\hline 
6 & $\left[\frac{k_{10}J_{10}}{\left(J_{10}+x_{6}\right)^{3}}-\frac{k_{9}x_{\mathrm{mpf}}J_{9}}{\left(J_{9}+1-x_{6}\right)^{3}}\right]\sigma_{66}$\tabularnewline
\hline 
7 & $-k'_{12}\sigma_{78}$\tabularnewline
\hline 
8 & $0$\tabularnewline
\hline
\end{tabular}
\end{table}
\begin{table}
\begin{centering}
\caption{\label{tab:diffusion-correct}Correction-term added to $B_{ii}(x)$
in \eqref{eq:2MA-sigmaii}.}
\smallskip{}

\par\end{centering}

\begin{tabular}{|c|c|}
\hline 
Index $i$ & $\varepsilon_{B_{ii}}(x,\sigma)=\frac{1}{2}\sum_{k,l}\frac{\partial^{2}B_{ii}}{\partial x_{k}\partial x_{l}}\sigma_{kl}\,$\tabularnewline
\hline
\hline 
1 & \selectlanguage{british}%
$k''_{2}\sigma_{13}+k'''_{2}\sigma_{15}$\selectlanguage{english}
\tabularnewline
\hline 
2 & $k''_{2}\sigma_{23}+k'''_{2}\sigma_{25}$\tabularnewline
\hline 
3 & $-\left[\frac{\left(k'_{4}x_{8}+k_{4}x_{\mathrm{mpf}}\right)J_{4}}{\left(J_{4}+x_{3}\right)^{3}}+\frac{\left(k'_{3}+k''_{3}x_{5}\right)J_{3}}{\left(J_{3}+1-x_{3}\right)^{3}}\right]\sigma_{33}-\frac{k''_{3}J_{3}\sigma_{35}}{\left(J_{3}+1-x_{3}\right)^{2}}+\frac{k'_{4}J_{4}\sigma_{38}}{\left(J_{4}+x_{3}\right)^{2}}$\tabularnewline
\hline 
4 & $0$\tabularnewline
\hline 
5 & $\frac{k_{7}J_{7}x_{6}\left(2\sigma_{45}-\sigma_{44}-\sigma_{55}\right)}{\left(J_{7}+x_{4}-x_{5}\right)^{3}}+\frac{k_{7}J_{7}\left(\sigma_{46}-\sigma_{56}\right)}{\left(J_{7}+x_{4}-x_{5}\right)^{2}}-\frac{k_{8}J_{8}}{\left(J_{8}+x_{5}\right)^{3}}\sigma_{55}$\tabularnewline
\hline 
6 & $-\left[\frac{k_{10}J_{10}}{\left(J_{10}+x_{6}\right)^{3}}+\frac{k_{9}x_{\mathrm{mpf}}J_{9}}{\left(J_{9}+1-x_{6}\right)^{3}}\right]\sigma_{66}$\tabularnewline
\hline 
7 & $k'_{12}\sigma_{78}$\tabularnewline
\hline 
8 & $0$\tabularnewline
\hline
\end{tabular}
\end{table}
\bibliographystyle{elsart-num}
\bibliography{man2ma}
\selectlanguage{english}

\end{document}